\documentclass[11pt,a4paper]{article}
\usepackage{jcappub}
\usepackage{amsfonts}
\usepackage{graphicx}
\expandafter\let\csname equation*\endcsname\relax 
\expandafter\let\csname endequation*\endcsname\relax 
\usepackage{amsmath}
\usepackage{amssymb}
\usepackage{bm}
\usepackage{mdframed}
\usepackage{mathtools}
\usepackage{slashed}
\usepackage{aas_macros}
\usepackage[colorlinks=true]{hyperref}

\newcommand{\pd}{\partial}

\renewcommand{\vec}[1]{\ensuremath{\mathbf{#1}}}

\title{Bosonization of Strong-Field Pair Plasma}

\author{Samuel E. Gralla}
\affiliation{Department of Physics, University of Arizona}

\emailAdd{sgralla@email.arizona.edu}

\abstract{A theory of magnetized pair plasma is obtained by the method of bosonization on the lowest Landau level.  Force-free electrodynamics emerges in the strong-field limit.  Potential applications to neutron star magnetospheres are suggested.}

\begin{document}
\maketitle 

\section{Introduction} 

Force-free electrodynamics (FFE) \cite{uchida1997general,komissarov2002,gralla-jacobson2014} consists of the following equations,
\begin{align}
F_{\mu \nu} \tilde{F}^{\mu \nu} = 0, \qquad F_{\mu \nu} F^{\mu \nu} > 0, \qquad \nabla_{[\mu}F_{\nu \rho]}=0, \qquad F_{\mu \nu} \nabla_\rho F^{\rho\mu} = 0,\label{FFE}
\end{align}
with $\tilde{F}_{\mu \nu}=(1/2) \epsilon_{\mu \nu \rho \sigma}F^{\rho \sigma}$.  From left to right, these are: degeneracy $(\mathbf{E} \cdot \mathbf{B}=0)$, magnetic domination $(\mathbf{B}^2>\mathbf{E}^2)$, no magnetic monopoles, and the \textit{force-free condition} $F_{\mu \nu}J^\nu=0$.  In general the current $J^\mu=\nabla_\nu F^{\mu \nu}$ is non-zero; this theory describes \textit{plasma}.  Interestingly, it describes plasma in many different regimes, from the lab \cite{gray-brown-dandurand2013} to the sun \cite{wiegelmann-sakurai-review2012} to neutron stars \cite{goldreich-julian1969} to black holes \cite{blandford-znajek1977}.  Recently it has been emphasized \cite{gralla-iqbal2018} that the differential equations in the theory (right two equations in \eqref{FFE}) are equivalent to the conservation of two currents,
\begin{align}
\nabla_\mu \tilde{F}^{\mu \nu}=0, \qquad \nabla_\mu T^{\mu \nu} = 0,
\end{align}
where $T_{\mu\nu}$ is the Maxwell stress-energy tensor.  This formulation helps explain why force-free fields are ubiquitous in nature: they follow from symmetries, independent of microscopics.

Nevertheless, it is illuminating and useful to have microscopic derivations valid for specific systems of interest.      One simple system is a collection of classical, non-interacting point charged particles (with both signs of charge present) immersed in a strong, magnetically dominated field of external origin.  As the particles respond to the field they will naturally cancel out its electric component, ultimately driving $\mathbf{E}\cdot\mathbf{B}$ entirely to zero if sufficiently plentiful.  However, the strength of the field means that individual charges are necessarily confined to field lines, executing small gyrations about a ``guiding center'' that slides freely along the line (e.g. \cite{cary-brizard2009}).  Mathematically, the Lorentz force law degenerates to the constraint $\mathbf{E} + \mathbf{v} \times \mathbf{B}=0$, which is written covariantly as $F_{\mu \nu} u^\mu=0$ for the four-velocity $u^\mu$.  (Formally, this is a zero-mass limit.)  If all charges move in this way, the total current $J^\mu$ similarly satisfies $F_{\mu \nu} J^\mu=0$, reproducing FFE.

This derivation is not particularly appropriate for the ultra-strong fields that occur near neutron stars, in the range of $10^8$ to $10^{15}$ Gauss.  In these fields the synchrotron cooling time is so short that any actual classical particle executing gyrations will quickly relax into the lowest Landau level of a flux quantum.  Some years ago, Thompson and Blaes \cite{thompson-blaes1998} provided a derivation more appropriate for these conditions.  They reasoned that the actual current is carried by the free momentum quantum number along a field line and hence is effectively two-dimensional.  Noting that the current in two-dimensions may be written $j^a = e \epsilon^{ab}\pd_b \Phi/(2\pi)$, they multiplied by the density of Landau levels $e B/(2\pi)$ to produce a four-dimensional current.  Finally they promoted this to a covariant expression assuming the fields vary slowly,
\begin{align}\label{JFFE}
J^\mu = -\frac{e^2}{8\pi^2} \epsilon^{\mu \nu \rho \sigma} \pd_\nu \Phi F_{\rho \sigma}.
\end{align}
The force-free condition $F_{\mu \nu} J^\mu=0$ then follows directly, assuming the fields are degenerate  ($F_{\mu \nu} \tilde{F}^{\mu \nu}=0$).\footnote{To see the equivalence, note that $F_{\mu \nu}\tilde{F}^{\mu \nu}=0$ implies that $F_{\mu \nu} = \alpha_{[\mu} \beta_{\nu]}$ for some $\alpha_\mu$ and $\beta_\mu$.} Even more interestingly, coupling the current \eqref{JFFE} to the Maxwell field provides an \textit{action principle} for FFE,\footnote{To reproduce the current in \eqref{JFFE} requires the coupling $(1/2) J^\mu A_\mu$, since $J^\mu$ depends on the field $A_\mu$.}
\begin{align}\label{SFFE}
S =  \int \sqrt{-g} d^4 x \left( -\frac{1}{4} F_{\mu \nu}F^{\mu \nu} - \frac{e^2}{16\pi^2} \Phi F_{\mu \nu} \tilde{F}^{\mu \nu} \right),
\end{align}
with $F_{\mu \nu}=2\nabla_{[\mu} A_{\nu]}$ as usual.  With this action principle the degeneracy constraint need not be put in, since $\Phi$ has become a  \textit{Lagrange multiplier} that enforces it automatically.  Varying with respect to $\Phi$ and $A_\mu$ produces
\begin{align}\label{PhiFFE}
\nabla_\nu F^{\mu \nu} = - \frac{e^2}{4\pi^2} \tilde{F}^{\mu \nu} \nabla_\nu \Phi, \qquad F_{\mu \nu}\tilde{F}^{\mu \nu} = 0.
\end{align}
Using the Maxwell equation $\nabla_\nu F^{\mu\nu}=J^\mu$ (here really the definition of $J^\mu$), the left equation is just \eqref{JFFE}, which was already remarked to be equivalent to the force-free condition $F_{\mu \nu}J^\mu=0$ provided $F_{\mu \nu}\tilde{F}^{\mu \nu}=0$.  The no-monopoles equation comes for free since $F=dA$, so one need only adjoin the inequality $F_{\mu \nu} F^{\mu \nu}>0$ (required for well-posed evolution \cite{komissarov2002,palenzuela-etal2011,pfeiffer-macfadyen2013,carrasco-reula2016}) to obtain the full force-free electrodynamics.   The formulation \eqref{PhiFFE} is also useful in practice since $\Phi$ can provide a conserved quantity along field lines \cite{gralla-lupsasca-philippov2017}.  Other actions for FFE are considered in \cite{uchida1997general,Glorioso:2018kcp,gralla-iqbal2018}.

Thompson and Blaes did not pursue the microscopics of their model too much further, but they did note that the field $\Phi$ ought to have a more fundamental description in terms of the method of \textit{bosonization} \cite{coleman1975,mandelstam1975}, wherein two-dimensional theories of fermions are transmuted into theories of bosons.  In this paper we will pursue the idea more fully and perform a top-down bosonization of a lowest-Landau-level QED plasma.  Aspects of our treatment will follow closely a similar calculation done in the context of holography \cite{blake-bolognesi-tong-wong2012}.  We will be led to a very peculiar regime of plasma physics---we term it \textit{coherent pair plasma}---that has not been considered previously.  The $\Phi$ field becomes fully dynamical, with a two-dimensional flux-weighted kinetic term accompanied by the famous sine-Gordon self-interaction.  We boldly treat this action classically.  Force-free electrodynamics emerges in a strong-field limit that we interpret as allowing free production of pairs by the Schwinger mechanism.\footnote{I am grateful to N. Iqbal for this observation.}

This can be regarded as a microscopic derivation of FFE, but ultimately we are most tempted by the phenomenology of this theory \textit{away} from the force-free limit.  Black hole and pulsar magnetospheres are believed to be connected with a variety of observational mysteries that FFE alone is unable to explain \cite{kotera-olinto-review2011,magnetoluminescence,katz-review2016,beskin-review2018}.  The theory we study here can support the unscreened electric fields $(\mathbf{E} \cdot \mathbf{B} \neq 0)$ needed to account for high-energy particles and radiation, and will likely have intricate features due to the sine-Gordon interaction.  We admit to have imagined the possibility that \textit{macroscopic} sine-Gordon solitons moving along field lines could provide the long-sought coherent radiation mechanism of pulsars \cite{beskin-review2018} and fast radio bursts \cite{katz-review2016}.

\section{Action for Coherent Pair Plasma}\label{sec:action}
 Fundamentally, a pair plasma is governed by Quantum Electrodynamics (QED),
\begin{align}\label{SQED}
S =\int d^4 x \left( \bar{\psi}\left( i \slashed{D} - m \right) \psi - \tfrac{1}{4} F_{\mu \nu} F^{\mu \nu} \right).
\end{align}
We set $\hbar=c=1$ and use the conventions of \cite{Peskin-Schroeder-book}.  We presume that a large scale $R\gg 1/m$ is set by boundary conditions (in the pulsar magnetosphere this would be the neutron star size), so that we can split the electromagnetic (EM) field into a coarse-grained piece and small-scale fluctuations.  We will neglect the fluctuations and treat the large-scale field classically. This eliminates many processes (such as Compton scattering and associated pair-production) that may be important in real pulsars, but we will see that an interesting theory survives.

We will work in a box of length $L_z$ and transverse area $A = L_x L_y$ over times $L_t$ such that $1/m\ll L_\mu \ll R$.  The EM field varies slowly over these scales and will be approximated as constant in space and time.  Assuming the field is not null (at least one of $F_{\mu \nu}F^{\mu \nu}$ and $F_{\mu \nu} \tilde{F}^{\mu \nu}$ not zero), by a suitable boost and rotation we may take $\vec{B}$ in the positive $z$ direction with $\vec{E}$ parallel, i.e. $\vec{E}=E_0 \bm{\hat{z}}, \vec{B}=B_0 \bm{\hat{z}}$ with $B_0>0$.  We represent the background field in the Landau gauge,
\begin{align}\label{landau}
A_\mu = (-z E_0, 0, x B_0, 0 ).
\end{align}
The field equation is now just the Dirac equation in this EM field,
\begin{align}\label{dirac}
\left( i \gamma^\mu \pd_\mu - e \gamma^\mu A_\mu - m\right) \psi = 0,
\end{align}
on which we will impose periodic boundary conditions at the box edges.

Without loss of generality we can solve the transverse dependence using separation of variables by
\begin{align}\label{psi-full}
\psi = \sum_{n=0}^{N} \sum_{j=0}^{\infty} \frac{e^{ -i k_n y}}{\sqrt{L_y}} \left( X_{n,j-1}(x) P_+ + X_{n,j}(x) P_- \right) \psi_{n,j}(t,z),
\end{align}
where $P_{\pm}=\tfrac{1}{2}(1\pm i \gamma^1 \gamma^2)$ projects onto spin-up $(+)$ and spin-down $(-)$ degrees of freedom.  The (orthonormal) transverse eigenfunctions $X_{n,j}(x)$ are given by
\begin{align}
X_{n,j} =\left(\frac{1}{2^j j! \sqrt{\pi} \ell_B }\right)^{\frac{1}{2}} \exp\left[ - \frac{1}{2} \left(\frac{x-x_n}{\ell_B}\right)^2 \right] H_j\left( \frac{x-x_n}{\ell_B} \right), \qquad j \geq 0,
\end{align} 
where $H_j$ are the Hermite polynomials, $X_{-1}\equiv 0$, and we introduce
\begin{align}
\ell_B = \frac{1}{\sqrt{eB_0}}, \qquad k_n=\frac{2\pi}{L_y} n, \qquad x_n = \frac{k_n}{eB_0}.
\end{align} 
This sum over $j$ is the decomposition into Landau levels, while the sum over $n$ reflects the degeneracy associated with the transverse size of the box.  In particular, the transverse momentum $k_n$ is discrete on account of the periodic boundary conditions, with range of summation arising from the requirement that the center $x_n$ of the wavefunction is within the box, i.e. $0<x_n<L_x$.  This gives $0 \leq n \leq e B_0 A/(2\pi)$, so that 
\begin{align}\label{N}
 N = \frac{e B_0 A}{2\pi}.
\end{align}

Plugging Eq.~\eqref{psi-full} back into the Dirac equation \eqref{dirac} yields equations for the longitudinal functions $\psi_{n,j}(t,z)$.  One can regard $P_\pm \psi_{n,j}$ as two-dimensional spinors labeled by three indicies $n,j,\pm$ and thereby determine a theory of coupled 2D spinors \cite{blake-bolognesi-tong-wong2012}, each representing a single spin degree of freedom in a single Landau level.  We will consider only the lowest Landau level $(j=0)$, which only has the spin-down degree of freedom.  In this case the ansatz \eqref{psi-full} becomes
 \begin{align}\label{psi-LLL}
\psi_{\rm LLL} = (\pi\ell_B^2)^{-\frac{1}{4}}  \sum_{n=0}^{N} \frac{e^{ -i k_n y}}{\sqrt{L_y}} e^{- \frac{1}{2} \left(\frac{x-x_n}{\ell_B}\right)^2} \xi_{n}(t,z), \qquad \xi_n = P_- \psi_{n,0}.
\end{align}
This restriction is mathematically consistent on account of our treatment of the EM field as classical---there is no coupling between the levels.  Physically, we require that any effects not included in our model (such as collisions with photons) transfer transverse ($xy$) momentum that is small compared to the typical energy spacing $eB_0/m$ of the Landau levels.\footnote{For very large fields $eB_0 \gg m^2$, the energy spacing is of order $\sqrt{e B_0} \ll e B_0/m$.}  We must also require the transverse particle density to be less than transverse density of states $eB_0/(2\pi)$, so that all particles can fit.

Plugging Eq.~\eqref{psi-LLL} into the Dirac equation \eqref{dirac} yields the 1+1-dimensional Dirac equation for each ``species'' $\xi_n$,
\begin{align}
\left( i\Gamma^M \pd_M - e \Gamma^M A_M - m\right)\xi_n = 0.
\end{align}
The capital latin index runs over only the $t,z$ components and $\Gamma^0=\gamma^0 P_-$ and $\Gamma^3=\gamma^3 P_-$ are 2D gamma-matrices.  For completeness use we also define the 2D chiral matrix $\Gamma^5=\gamma^5 P_-=-\Gamma^0 \Gamma^3$.  In the Weyl representation, these are related to the Paul matrices by $\Gamma^0=\sigma^1, \Gamma^3=-i\sigma^2, \Gamma^5=-\sigma^3$. The action for this theory is just
\begin{align}\label{SLLL}
S_{\xi} =  \int dt dz \sum_{n=0}^{N} \bar{\xi}_n \left( i \Gamma^M  \left(\pd_M + i e A_M\right) - m \right) \xi_n,
\end{align}
which can also be derived by plugging \eqref{psi-LLL} into the Dirac action and performing the integral over the transverse directions.  We have chosen the normalizations so \eqref{SLLL} arises directly with no extra prefactor.

We now employ the method of bosonization \cite{coleman1975,mandelstam1975}.  %As Eq.~\eqref{SLLL} is a \textit{free} theory of $N$ 1+1-dimensional charged fermions in a uniform background electric field, we may use the standard map to bosonic degress of freedom \cite{coleman1975,mandelstam1975}.  
The relations between the boson fields $\phi_n$ and the fermions $\xi_n$ is non-local and can depend on the theory being bosonized, but the fermion bilinears have the simple substitution rules (e.g. \cite{shankar-book}) (no sum on $n$)%\footnote{\SG{We are ignoring the so-called Klein factors to ensure that different species of Fermions properly anti-commute.  This will be irrelevant in the coherent limit we take.}}
\begin{subequations}\label{bilinears}
\begin{align}
i \bar{\xi}_n \Gamma^A \pd_A \xi_n &\leftrightarrow \frac{1}{8\pi} \pd^A \phi_n \pd_A \phi_n \\
\bar{\xi}_n \Gamma^A \xi_n & \leftrightarrow \frac{1}{2\pi} \epsilon^{AB} \pd_B \phi_n \label{Jel} \\
\bar{\xi}_n \Gamma^5 \Gamma^A \xi_n & \leftrightarrow \frac{1}{2\pi} \pd^A \phi_n \label{Jax} \\ 
\bar{\xi}_n \xi_n & \leftrightarrow - \frac{\Lambda}{4\pi} \cos \phi_n.
\end{align}
\end{subequations}
Here $\Lambda$ is a convention-dependent mass scale used to define composite operators; it must be matched to the analogous scale used on the fermion side in comparing any given calculation.
%Here $\Lambda$ is a mass scale to be matched to a suitable mass on the fermion side in comparing any given calculation.  %\footnote{The arbitrary mass scale is used to define composite operators such as $\cos \phi_n$, with the details depending on the theory being bosonized and the particular approach to bosonization.  These details are not important for the present analysis.}
Making these substitutions, the action \eqref{SLLL} becomes 
\begin{align}\label{SLLLPhi}
S_{\phi} =  \int dt dz  \sum_{n=0}^{N}\left( \frac{1}{8\pi} \pd^A \phi_n \pd_A \phi_n - \frac{m \Lambda}{4\pi} \left(1-\cos \phi_n\right) - \frac{ e E_0}{2\pi} \phi_n \right) ,
\end{align}
where we have added a constant to make the potential equal zero at its minima.  We have also integrated by parts on the last term.

%Under the boson/fermion duality, states of definite fermion number are coherent boson states, while boson excitations correspond to electron-positron pairs that do not separate, effectively a new meson in the theory \cite{bosonization-for-beginners}.

One utility in writing a theory in a new way is that it may suggest interesting new limits.  Our strategy will be to take the classical limit of the bosonic action, i.e., to consider its equations of motion for commuting classical fields $\phi_n$.  However,  the arbitrary scale $\Lambda$ appears in the action, so we must assign it a value for this limit.  The simplest argument is to set $\Lambda\sim m$ to make the mass of the sine-Gordon soliton of order the fermion mass.  By Eq.~\eqref{Jel} the fermion electric charge becomes a topological charge of the boson theory, and the soliton contains precisely one unit (e.g. \cite{weinberg-book}); it is therefore naturally identified with the fermion.  Additional evidence for the physical relevance of the bosonic classical limit arises from the mapping of states in the duality, where one-fermion states correspond to coherent states of the boson \cite{bosonization-for-beginners}, which are in some sense ``most classical''.  
%See also \cite{hirata-minakata-1986} for further discussion of the semi-classical approximation of 
%This is supported by various  calculations (e.g. \cite{hirata-minakata-1986,weinberg-book}) suggesting that classical solutions of the sine-Gordon equations capture some essential features of the quantum theory, with the solitonic solutions representing the elementary fermion.   Further evidence arises from the mapping of states in the duality, where one-fermion states correspond to coherent states of the boson \cite{bosonization-for-beginners}.  
Further discussion may be found in Ref.~\cite{blake-bolognesi-tong-wong2012}, where it is concluded that in general $\Lambda$ is to be identified with a physical scale in the relevant semi-classical limit.  The only scale here is the renormalized mass $m_R(B_0)$.  However, since the mass runs only weakly with magnetic field \cite{janovici1969,constantinescu1972}, we will just consider it constant.  We thus write 
\begin{align}\label{Lambda}
\Lambda m=\bar{m}^2,
\end{align}
 where $\bar{m}$ is of order the electron mass.  Most conservatively, we could simply regard $\bar{m}$ as a free parameter in the theory.

%The meaning of this step is most clear in the special case $m=0$, corresponding to ultrarelativistic fermions.  In this case the bosonic theory is free and the classical limit can be interpreted as high occupation number of small bosonic excitations above the vacuum.  On the fermion side, these excitations may be interpreted as  electron-positron pairs that move at light speed and hence do not separate \cite{coleman1975}.  \SG{wait... don't they carry charge???} Thus the $m=0$ bosonic classical limit corresponds to a quasi-neutral ultrarelativistic plasma.  \SG{hmm... do the EOM give quasi-neutral?  I don't think so...}

Treating each $\phi_n$ as a classical field, our next step is to relate the $\phi_n$ to eachother.  Each $\phi_n$ can be regarded as living on a magnetic field line that represents one quantum of magnetic flux.  We will assume that the field configuration does not vary too much from field line to field line, so that the plasma is \textit{coherent} over scales of order $\ell_B$.  Formally we introduce a transverse coherence length $\zeta_\perp$ representing the scale over which the $\phi_n$ agree with eachother.  We say the plasma is coherent if
\begin{align}
\textrm{coherent plasma:}\qquad \zeta_\perp \gg \ell_B. \label{coherence-length} 
\end{align}
Taking the transverse box size of order the coherence length ($L_x \sim L_y \sim \zeta_\perp)$, we may approximate the $\phi_n$ as sharing a single field configuration $\Phi$,
\begin{align}\label{manyasone}
\phi_1 = \phi_2 = \dots = \phi_N \equiv \Phi.
\end{align}
%As each $\phi_n$ is bosonic, the classical limit is sensible when a given set of states is occupied by many particles (corresponding to the neutral e+/e- mesons). However, we can also expect the classical limit to work even for states of definite electron number, since these are coherent on the boson side.  A separate notion of coherence that we advance is the assumption \eqref{manyasone} that the boson fields are all in the same configuration.  We will characterize the validity of this assumption by introducing 
%The further assumption \eqref{manyasone} will be valid if the mechanism creating the plasma operates on scales much larger than $\ell_B$.  To be precise, we can assign 
%a transverse coherence length $\zeta_\perp$ representing the scale over which the $\phi_n$ agree.  We say the plasma is coherent if
%\begin{align}
%\textrm{coherent plasma:}\qquad \zeta_\perp \gg \ell_B, \label{coherence-length} 
%\end{align}
%in which case $\Phi$ well-describes the plasma and varies on transverse scales $\zeta_\perp$. 
One can expect such coherence to emerge when the mechanism creating the plasma operates on scales much larger than $\ell_B$.  In pulsars, the ultimate driver of plasma production is the rotating neutron star (see discussion in Sec.~\ref{sec:applications} below), so the natural scale is kilometers.   The magnetic length $\ell_B$ is vastly smaller ($\ell_B \approx 10^{-12}/\sqrt{B_{12}}$ meters, where $B_{12}$ is the magnetic field in units of $10^{12}$ Gauss), and we expect no difficulty with coherence.

Plugging Eqs.~\eqref{Lambda} and \eqref{manyasone} into the action \eqref{SLLLPhi} and using $N=e B_0 A/(2\pi)$ [Eq.~\eqref{N}], we find
\begin{align}
S^{\rm \Phi}_{\rm 1+1} & = \int dt dz  \frac{e B_0 A}{2\pi} \left(\frac{1}{8\pi} \pd^A \Phi \pd_A \Phi - \frac{\bar{m}^2}{4 \pi} \left(1-\cos \Phi\right) - \frac{e}{2\pi} E_0 \Phi \right) \label{Scoherent0} \\
& =  \int d^4x \frac{e B_0}{2\pi} \left(\frac{1}{8\pi} \pd^A \Phi \pd_A \Phi - \frac{\bar{m}^2}{4 \pi} \left(1-\cos \Phi\right) - \frac{e}{2\pi} E_0 \Phi \right).\label{Scoherent}
\end{align}
In the second step we have noted that the transverse area $A$ of the box is just $\int dx dy$.  

Eq.~\eqref{Scoherent} gives the dynamics of the plasma in a local Lorentz frame where the electric and magnetic fields are constant and parallel, with $\Phi$ independent of $x$ and $y$.  Requiring that this action emerge in every such local frame gives rise, via considerations of covariance, to a global action for a field $\Phi$ now allowed to depend on all four spacetime coordinates.  To find this action we introduce a covariant expression for a projector to the 1+1D subspace\footnote{\label{hi}Any non-null electromagnetic field (at least one of $F_{\mu \nu}F^{\mu \nu}$ and $F_{\mu \nu} \tilde{F}^{\mu \nu}$ not zero) defines a preferred family of frames where $\mathbf{B}$ and $\mathbf{E}$ are parallel (including the case where one vanishes).  The frames are related by Lorentz boosts along the field direction (call it $\hat{\bm{z}}$), which preserve the fields $\mathbf{E}=E_0 \hat{\bm{z}}$ and $\mathbf{B}=B_0 \hat{\bm{z}}$.  The projector $h_{\mu \nu}$ has components $(1,0,0,-1)$ in any such frame and acts as the two-dimensional metric for the theory \eqref{SLLLPhi}.}
\begin{align}\label{h}
h_{\mu \nu} = \frac{F_{\mu \alpha} F^{\alpha}{}_{\nu} + B_0^2 g_{\mu \nu}}{E_0^2 + B_0^2}.
\end{align}
We then relate $E_0$ and $B_0$ to the tensorially natural invariants $a$ and $b$ (organized as in  \cite{jacobson2015}),
\begin{align}
B_0^2 - E_0^2 & = \vec{B}^2 - \vec{E}^2 = \tfrac{1}{2} F_{\mu \nu} F^{\mu \nu} \equiv a \\
2 E_0 B_0 & = 2 \vec{E} \cdot \vec{B} = \tfrac{1}{2} F^{\mu \nu} \tilde{F}_{\mu \nu} \equiv b.
\end{align}
Noting $\sqrt{a^2+b^2}=E_0^2+B_0^2$, in particular we have
\begin{align}
\sqrt{2} E_0 & = \textrm{sign}[b] \sqrt{ \sqrt{a^2+b^2} - a} \\
\sqrt{2} B_0 & = \sqrt{ \sqrt{a^2+b^2} + a}. \label{B0}
\end{align}
Finally, we will promote to curved spacetime by $\pd \to \nabla$ as usual.  Using these ingredients to express \eqref{Scoherent} covariantly, allowing all fields to depend on all coordinates, and re-introducing the $F^2$ term from \eqref{SQED} gives the complete action for a coherent plasma as\footnote{Recall that in writing Eq.~\eqref{SLLLPhi} we made a particular choice of the overall constant in the dimensionally-reduced action.  Had we made a different choice, a term proportional to $B_0$ would appear in the final action \eqref{Sfinal} even as $\Phi \to 0$, and the theory would not properly reduce to free Maxwell theory.}
\begin{align}\label{Sfinal}
S =  \int \sqrt{-g} d^4 x \left( -\frac{1}{4} F_{\mu \nu}F^{\mu \nu} -  \frac{e^2}{16\pi^2} \Phi \tilde{F}^{\mu \nu} F_{\mu \nu} + \frac{e B_0}{8\pi^2} \left( \tfrac{1}{2} h^{\mu \nu} \nabla_\mu \Phi \nabla_\nu\Phi - \bar{m}^2(1-\cos \Phi)\right)  \right).
\end{align}
The degrees of freedom are $A_\mu$ and $\Phi$, with $B_0$ and $h_{\mu \nu}$ are constructed from $F_{\mu \nu}=2 \nabla_{[\mu} A_{\nu]}$ using Eqs.~\eqref{h}-\eqref{B0}.  Notice that, despite the axionic coupling $\Phi \tilde{F}^{\mu \nu} F_{\mu \nu}$, the field $\Phi$ is not an axion: its kinetic term is two-dimensional (projected to the field direction) and weighted by the local field $B_0$.

Before proceeding, we wish to emphasize that the theory \eqref{Sfinal} contains no additional tensor structure beyond the Maxwell field.  In particular, there is no preferred ``plasma frame'' (four-velocity $u^\mu$), and frame-dependent concepts like temperature and number density do not appear in the theory or in its conditions for validity (listed during the derivation).  Only the preferred \textit{family} of frames defined by the Maxwell field (see footnote \ref{hi}) can be involved in any physics related to this theory.   This manifests as Lorentz-invariance in the dimensionally reduced theory \eqref{Scoherent0}, and is the essential reason why the conditions for validity involve only transverse directions.

However, it is instructive to relate to other descriptions of plasma that do involve a preferred frame.  For example, a temperature $T$ in some frame $u^\mu$ corresponds to transverse momenta at most of order $\gamma_\perp T$ in the preferred family of frames, where $\gamma_\perp=\sqrt{h_{\mu \nu}u^\mu u^\nu}$ is the minimal Lorentz factor relative to the preferred family (we set Boltzmann's constant to unity).  Thus we require $T \ll  e B_0/(\gamma_\perp m)$ for particles to remain in the lowest Landau level.  Similarly, we require the number densities $n_{\pm}$ of positrons and electrons to satisfy $n_{\pm}^{2/3} \ll e B_0/(2\pi \gamma_\perp)$ so all particles fit.  However, we avoid saying that \eqref{Sfinal} represents a low-temperature, low-density plasma; instead, it corresponds to a strong-field regime where temperature and density are simply not relevant concepts.

Finally, we discuss the interpretation of $\Phi$, which is the only degree of freedom describing the plasma.  This field emerged from treating the bosonized degrees of freedom classically and insisting on slow variation from field line to field line.  Its derivatives encode the electric  \eqref{Jel} and axial \eqref{Jax} current flowing along the local field direction.  Because of the use of the duality, its interpretation in terms of electrons and positrons is unclear.  However, this is precisely the point: by using a new description of the plasma, we hope to reveal properties that are obscure in the traditional language.  
%That said, since plasma production processes are generally understood in the langauge of fermions, a better understanding of the correspondence is necessary to determine the mechanism by which a plasma in the regime \eqref{Sfinal} could be created.  %Note, however, that plasma production processes are generally understood in the language of fermions, so a better understanding of the correspondence is necessary to determine what mechanism could place plasma in the regime described by \eqref{Sfinal}.

 %However, the Maxwell field does define a \textit{family} of frames where the electric and magnetic fields are parallel, which were used extensively in the derivation of \eqref{Sfinal}.  These frames are related by boosts along the field direction, which preserve $s$ and $B_0$ and manifest as the boost invariance of the dimensionally-reduced theory \eqref{Scoherent}.  The conditions for validity (listed during the derivation) .  

\section{Force-free Limit}\label{sec:force-free}

Force-free electrodynamics arises if we can drop the last two terms in \eqref{Sfinal} in order to reproduce the Thompson-Blaes action \eqref{SFFE}.  We now discuss two circumstances in which this is possible, and mention a third regime where it may also make sense.

The first is if $\Phi$ is globally near a vacuum $\Phi_0 =2\pi n$, where $n$ is an integer.  Using $\delta \Phi = \Phi - \Phi_0 \ll 1$, the leading appearance of $\delta \Phi$ in \eqref{Sfinal} is the second term; the last two terms are $O(\delta \Phi^2)$ and may be neglected.  This reproduces the action \eqref{SFFE} with $\Phi$ replaced by $\delta \Phi$.  Varying with respect to $\delta \Phi$ still produces the non-linear equations of FFE in terms of $\delta \Phi$, but only solutions with $\delta \Phi \ll 1$ should be considered.  In light of the formula \eqref{JFFE} for the force-free current, this translates physically to the assumption 
\begin{align}
|J| \ll B_0/\mathcal{R},
\end{align}
where $\mathcal{R}$ is a typical scale of variation.  Thus the $\delta \Phi \ll 1$ sector should properly only include solutions with nearly-zero charge-current; that is, only \textit{linearized} solutions about current-free backgrounds.  %This is an interesting class of solutions (describing, for example, energy extraction from slowly spinning black holes \cite{blandford-znajek1977}), but it is not the most general case.  Physically, it describes a nearly-neutral plasma.
This corresponds to a nearly neutral plasma. 

Full non-linear FFE emerges in a separate limit where the fields are taken to be very \textit{strong}.  To illustrate, we repeat \eqref{Sfinal} and annotate the size of each term,
\begin{align}
S & =  \int \sqrt{-g} d^4 x \left[ -\frac{1}{4} F_{\mu \nu}F^{\mu \nu} - \frac{e^2}{16\pi^2} \Phi \tilde{F}^{\mu \nu} F_{\mu \nu} + \frac{e B_0}{8\pi^2} \left( \tfrac{1}{2} h^{\mu \nu} \nabla_\mu \Phi \nabla_\nu\Phi - \bar{m}^2(1-\cos \Phi)\right)  \right] \nonumber \\
& \qquad \qquad \! \! \sim  B^2 \left[ \quad \ \ 1 \qquad+ \qquad \alpha  \qquad \  + \quad \  \alpha \frac{B_c}{B} \left( \quad \frac{1}{(m \zeta_{||})^2} \quad  \ \ + \qquad \ \ 1 \qquad \right)\right]\label{scales} 
\end{align}
Here $B$ is a typical field scale (we do not distinguish between electric and magnetic), $\alpha=e^2$ is the fine structure constant, and $B_c$ is the QED-critical field strength,
\begin{align}
B_c=m^2/e \approx 4 \times 10^{14} \textrm{ Gauss}.
\end{align}
The scalar $\Phi$ is assumed order-one and varying on scales $\zeta_\parallel$ and $\zeta_\perp$ parallel and perpendicular to the local field direction.\footnote{The perpendicular scale $\zeta_\perp$ is defined using the family of frames as in Sec.~\ref{sec:action}.  For the parallel scale $\zeta_\parallel$ there is no invariant definition and we imagine comparing the size of terms in the action relative to a preferred frame given externally by boundary conditions (e.g. the frame of a neutron star).}  Since all derivatives are projected onto the field direction, only $\zeta_\parallel$ appears in \eqref{scales}.   We see that the last two terms disappear at strong field,
\begin{align}
\textrm{Force-free limit: } \qquad B \gg B_c.
\end{align}

This means that force-free behavior emerges as the field exceeds the quantum critical scale.  We interpret  this on the fermion side of the duality as due to the free availability of pairs via the Schwinger mechanism; these pairs can screen $E_0$ to zero, as required in FFE.  (Below the critical scale, no other pair production mechanisms are available as our EM field is classical.)  It is not clear exactly how to interpret Schwinger pair-production on the boson side, but it clearly must occur as it is a one-loop process involving only a background electromagnetic field. The hidden role of loops can be seen from the appearance of $\alpha$ in the bosonized terms of the action \eqref{scales}.  This is characteristic of bosonization: it ``classicalizes'' one-loop fermion effects \cite{hirata-minakata-1986,chiral-magnetic-wave,blake-bolognesi-tong-wong2012}.

We must also check that the force-free limit preserves the assumptions of the model.  The transverse coherence assumption \eqref{coherence-length} may be written
\begin{align}\label{B-coherent}
\textrm{coherence: } \qquad m \zeta_\perp \gg \sqrt{\frac{B_c}{B}}.
\end{align}
Thus as $B \to \infty$ it becomes easier to maintain coherence: the force-free limit is consistent.  

By contrast, coherence is impossible to maintain as $B \to 0$.  However, for macroscopic coherence lengths ($\zeta_\perp m \gg 1$) there is a wide range of field strengths where coherence is possible at weaker fields,
\begin{align}\label{M-coherent}
\frac{B_c}{(m \zeta_\perp)^2} \ll B \ll B_c.
\end{align}
In this regime the last two terms in the action \eqref{scales} are dominant, suggesting that the sine-Gordon dynamics on each field line decouples from the rest of the plasma.  We may then imagine current moving only on field lines, either in smooth or solitonic field configurations.  By some kind of coarse-graining over the microscopics, it may be possible to replace $\Phi$ with a simpler model of current flowing along the field lines.  One could then invoke a sufficiency of charge to screen the electric field ($\mathbf{E} \cdot \mathbf{B} \to 0$).  The dynamics of this particular class of weak-field solution are then force-free, since the condition $F_{\mu \nu}J^\mu=0$ is equivalent to the statement that the charge-current points along the field lines of a degenerate electromagnetic field \cite{gralla-jacobson2014}.  It would be very interesting to make this coarse-graining procedure precise.

\section{Discussion}\label{sec:applications}

The main result of this paper is the action \eqref{Sfinal} for a ``coherent pair plasma''.  Beyond its theoretical interest, we propose neutron star magnetospheres as a candidate physical application.  It is generally believed that pulsars shine because of the presence of plasma outside the neutron star (e.g. \cite{beskin-review2018}).  Even if no plasma is present at birth, the enormous voltage (some ten orders of magnitude greater than $m$) generated by the rotation of the conducting neutron star crust through the stellar magnetic field will ``spark'' the vacuum in one way or another to allow a current to flow through space.  The most likely process \cite{sturrock1971} is the acceleration of stray charges along curved magnetic field lines, which radiate gamma-ray photons that pair-produce in the strong magnetic field, leading to a pair-creation cascade.  Other processes may occur; the main point is that energetics seem to make pair production unavoidable over a wide swath of parameter space that corresponds to observed pulsars.

That plasma production occurs is thus reasonably certain, but the plasma density at which it \textit{shuts off} is less so.  The non-zero $\mathbf{E} \cdot \mathbf{B}$ drives the pair production, but the pairs begin to screen this field as they become more plentiful.  Pair-production will stop once $\mathbf{E} \cdot \mathbf{B}$ becomes sufficiently small, but exactly \textit{how} small depends on the microscopic model.  The action \eqref{Sfinal} seems most useful in a \textit{charge-starved} regime where there is not enough plasma to completely screen the field.  Usually this regime is explored with phenomenological resistivity \cite{gruzinov2008,kalapotharakos-kazanas-harding2012,li-spitkovsky-tchekhovskoy2012} or particle-in-cell simulations (see \cite{chen-beloborodov2014,philippov-spitkovsky-cerutti2015} and subsequent references).  Here we suggest an alternative approach that is in a sense hybrid, since plasma is described by a smooth field $\Phi$ that can nonetheless develop particle-like features on account of the sine-Gordon term with mass scale $\bar{m}$.

However, this separation of scales will also make the field equations rather difficult to solve numerically.  Furthermore, it is connected with a lingering theoretical worry about the consistency of the theory: We assumed that the EM field varies on scales much larger than $1/m$, but obtained a theory where it is coupled to the scale $\bar{m} \sim m$.  Does this mean the EM field will always develop structure on small scales, invalidating the original assumption?  We think not, on account of the scale appearing as a self-interaction only.  Barring adverse initial conditions, we expect fine structure to develop mainly in longitudinal directions of the $\Phi$ field, with significant influence on the electromagnetic field only on much larger scales.  Of course, generic solutions will presumably show \textit{some} small-scale variation of $F_{\mu \nu}$, and in applications it may be necessary to coarse-grain this deviation away.  One could imagine beating the separation of scales using an effective field theory framework, such as that recently proposed near the force-free limit \cite{gralla-iqbal2018}.  These difficulties would also disappear in an ultrarelativistic limit where the mass term could be dropped.

We now discuss some future directions.  First, it would be interesting to study the bosonization at the level of mapping individual fields, instead of relying in substitution rules \eqref{bilinears}.  This would improve the rigor of the analysis and clarify the meaning of the scale $\bar{m}$.  Second, it would be interesting to include some aspects of the quantized electromagnetic field and/or go beyond the lowest Landau level approximation.  It would be illuminating to see whether force-free dynamics emerge naturally at sub-critical field strengths if a second pair-production mechanism is enabled.  Third, it would be interesting to relax the assumption \eqref{manyasone} of complete coherence between the degenerate Landau levels.  In principle there should be a perturbative expansion away from true coherence, or an alternative framework where the field $\Phi$ becomes a statistical average.  This is likely related to the coarse graining issues discussed above.

Fourth, it would be very interesting to explore connections with the ``chiral magnetic wave'' expected in quark-gluon plasma, which has also been studied with bosonization on the lowest Landau level \cite{chiral-magnetic-wave}.  In fact, the chiral magnetic effect \cite{fukushima-kharzeev-warringa2008} itself is a kind of force-free dynamics (since the current flows along the magnetic field), and effective actions with the axionic coupling $\Phi F_{\mu \nu} \tilde{F}^{\mu \nu}$ indeed appear in its study \cite{kharzeev-2010,kharzeev-review2014}.  The pulsar magnetosphere can be understood as a \textit{reverse} chiral magnetic effect: Instead of a microscopic chiral imbalance becoming a battery to drive current along the field, a macroscopic battery (the rotating neutron star) drives current in the lowest Landau level that results that results (at least in the massless limit) in a net flow of chirality.  It is not clear whether there are any observational consequences of this chirality of the pulsar magnetosphere.

Finally, it would be very interesting to attempt direct numerical simulation of the field equations associated with \eqref{Sfinal}.  For inspiration, we conclude by writing them out:
\begin{align}
\nabla_\nu F^{\mu \nu} & = -\frac{e^2}{4\pi^2} \tilde{F}^{\mu \nu} \nabla_\nu   \Phi + \frac{e}{8\pi^2} \nabla_\nu \mathcal{J}^{\mu \nu} \\
\nabla_\mu(B_0 h^{\mu \nu} \nabla_\nu \Phi) + \bar{m}^2 B_0 \sin \Phi & = - \tfrac{1}{2} e \tilde{F}^{\mu \nu} F_{\mu \nu},
\end{align}
where the antisymmetric tensor $\mathcal{J}^{\mu \nu}$ is given by
\begin{align}
\mathcal{J}^{\mu \nu} & = \frac{F^{\mu \nu}+H^{\mu \nu}}{2B_0} \Big[ \tfrac{1}{2} h^{\rho \sigma}\nabla_\rho \Phi \nabla_\sigma \Phi - \bar{m}^2(1-\cos \Phi)\Big] \nonumber \\
& \qquad + \frac{B_0}{E_0^2+B_0^2} \Big[2 g_{\rho}{}^{[\mu} F^{\nu]}{}_\sigma + \tfrac{1}{2} F^{\mu \nu} g_{\rho \sigma} + H^{\mu \nu}(\tfrac{1}{2} g_{\rho \sigma} - h_{\rho \sigma}) \Big] \nabla^\rho \Phi \nabla^\sigma \Phi,
\end{align}
with
\begin{align}
H^{\mu \nu} = \frac{a F^{\mu \nu} + b \tilde{F}^{\mu \nu}}{E_0^2+B_0^2}.
\end{align}
In these expressions, $E_0$, $B_0$, $a$, $b$ and $h_{\mu \nu}$ are constructed from $F_{\mu \nu}$ using Eqs.~\eqref{h}-\eqref{B0}.

\section*{Acknowledgments}
It is a pleasure to acknowledge Marat Freytsis, Nabil Iqbal, Alex Lupsasca, Sanjay Reddy, and Srimoyee Sen for many helpful conversations.  I am also grateful to the Aspen Center for Physics (which is supported by National Science Foundation grant PHY-1607611), where some of the ideas were formulated. This work was supported in part by the NSF under award PHY-1752809 to the University of Arizona. 

%\section*{References}
\bibliography{bosonization}

\end{document}